\documentclass[a4paper]{article}

\usepackage{icrc2013}
\usepackage[english]{babel}

\newcommand\beq{\begin{equation}}
\newcommand\beql[1]{\begin{equation} \label{#1}}
\newcommand\eeq{\end{equation}}
\newcommand\ben{\begin{eqnarray}}
\newcommand\een{\end{eqnarray}}
\newcommand\bea{\begin{array}}
\newcommand\eea{\end{array}}
\newcommand\bem{\begin{displaymath}}
\newcommand\eem{\end{displaymath}}

\newcommand\eqa[1]{Eq.(\ref{#1})}
\newcommand\eqb[1]{Eqs.(\ref{#1})}
\newcommand\eqc[1]{(\ref{#1})}

\newcommand\fig[1]{Fig.\ref{#1}}
\newcommand\figs[1]{Figs.\ref{#1}}
\newcommand\figg[1]{\ref{#1}}

\newcommand\app[1]{Appendix~\ref{#1}}

\newcommand\qqb{\qquad}

\newcommand\Ssum[2]{\sum \limits_{#1}^{#2}}

\newcommand\sgn{\rm sgn}

\newcommand\Non{N_{\rm on} }

\newcommand\Noff{N_{\rm off} }

\newcommand\ton{t_{\rm on} }
\newcommand\toff{t_{\rm off} }
\newcommand\won{w_{\rm on} }
\newcommand\woff{w_{\rm off} }
\newcommand\mon{\mu_{\rm on} }
\newcommand\moff{\mu_{\rm off} }
\newcommand\mono{\mu_{\rm on,0} }
\newcommand\moffo{\mu_{\rm off,0} }
\newcommand\lon{\lambda_{\rm on} }
\newcommand\loff{\lambda_{\rm off} }
\newcommand\Xon{X_{\rm on} }
\newcommand\Xoff{X_{\rm off} }

\newcommand\ms{\mu_{\rm s} }
\newcommand\mb{\mu_{\rm b} }
\newcommand\mso{\mu_{\rm s,0} }
\newcommand\mbo{\mu_{\rm b,0} }

\newcommand\pex{p_{\rm e} }
\newcommand\pde{p_{\rm d} }
\newcommand\Bii{{\rm Bi} }
\newcommand\SBii{S_{\rm Bi} }
\newcommand\SLM{S_{\rm LM} }
\newcommand\Noo{{\rm N} }
\newcommand\Poo{{\rm Po} }

\title{Testing time variability of gamma-ray flux}

\shorttitle{Testing time variability}

\authors{
Dalibor Nosek$^{1}$,
Stanislav Stefanik$^{1}$,
Jana Noskova$^{2}$.
}

\afiliations{
$^1$ Institute of Particle and Nuclear Physics, 
Faculty of Mathematics and Physics, Charles University, 
Prague, Czech Republic \\
$^2$ Department of Mathematics, 
Faculty of Civil Engineering, Czech Technical University,
Prague, Czech Republic
}

\email{stefanik\_ stanislav@yahoo.com}

\abstract{
A way of examining a hypothetical non--zero $\gamma$--ray signal 
for the time changes is presented. 
The time variability of the recently observed $\gamma$--ray 
source PKS~2155--304 is discussed. 
Several measurements were found to be excessive or deficient 
with large significances on time scales of months and days. 
}


\begin{document}

\maketitle

\section{Introduction}
\label{Sec01}

The asymptotic Li--Ma technique~\cite{Lim01} is traditionally used 
in $\gamma$--astronomy to confirm positive results in searching for 
discrete sources.
In high energy physics, different signal--to--background methods or 
more advanced techniques based, for example, on Bayesian reasoning 
are applied to distinguish signal from background~\cite{Cou01}.
To our knowledge, no simple procedure aiming at testing the change 
in a beforehand proven source activity has been established on 
statistical grounds, however.

In the following, we describe briefly the on--off problem and introduce
two different statistical measures for refusing no--source or no--sink 
hypotheses, binomial and Li--Ma significances.
As a main goal, we present a modified on--off method to test the change 
of a given non--zero source activity.
This method is applied to demonstrate the time variability of 
the $\gamma$--ray flux observed in the direction of the source 
PKS~2155--304 on different time scales~\cite{Hess01,Hess02}.

\section{The on-off method}
\label{Sec02}

The on--off analysis is widely used in many branches of physics,
especially in $\gamma$--ray astronomy or in high energy physics, 
see e.g. Refs.~\cite{Lim01,Cou01}.
In $\gamma$--ray astronomy, the on--off problem is communicated in 
the following way.
Consider an observation of events in a given time aiming to detect 
their source in the on--source region on the sky the choice of which 
is motivated by previous observations or by other arguments.
Their on--source excess is judged by comparing the number of events
with arrival directions pointing to the on--source region, $\Non$, 
to the number of events with arrival directions in a control 
off--source region, $\Noff$, where no signal events are expected.
Both these observations are treated as Poissonian random variables.
The expected ratio of the count numbers in these regions is assumed 
to be known provided no source is present in the on--source region.
This ratio, the on--off parameter, is simply 
$\alpha = \frac{\won}{\woff}$ where
$\won$ and $\woff$ are given weights of the on-- and off-source regions, 
respectively.\footnote{
Usually, a time process of observation where events occur continuously 
and independently of one another in time is considered.
In such a case, the weights are simply given by observational times,
$\won = \ton$ and $\woff = \toff$.
One can also deal with the spatial process. 
It is, for example, introduced as a spatial dependence of 
exposure--weighted areas enclosed in circles with radii given 
by elements of an ordered set of separation angles of events 
as measured from a source direction.
Then, the weights are given by these exposure--weighted areas.
}
In this sense, the on--off problem consists in constructing 
a hypothesis test for the ratio of two unknown Poissonian 
parameters, $\mon$ and $\moff$, assumed to give observed on-- 
and off--source counts, respectively.
The null hypothesis of no source in the on--source region, 
expressed by the equality 
$\mon = \alpha \moff$, 
is then tested against a one--sided alternative of an excess 
of on--source counts,
i.e. $\mon > \alpha \moff$.
This test is easily modified to judge a deficit of events 
in a region where their sink is expected.

Let us assume in the following that the source is ascertained with 
a given activity such that $\mon = \alpha \beta \moff$ where 
a source parameter $\beta$ known in advance describes the excess 
of on--source counts above off--source ones.
Then a question arises whether follow--up counts do agree with 
previous observations or not.
In this case, we propose a test of the null hypothesis 
expressed by $\mon = \beta \alpha \moff$ against a two--sided 
alternative of an excess or deficit of on--source counts,
i.e. $\mon > \beta \alpha \moff$ or $\mon < \beta \alpha \moff$. 

A suggested method is described in \app{App01}. 
A simple binomial treatment of the on--off problem relying on 
the aforementioned non--zero source conjecture is described 
in \app{App01a}.
A modified asymptotic Li--Ma significance that may be used 
to express deviation of observed counts from a given non--zero 
source is introduced in \app{App01b}.
We show that within a classical statistical treatment reasonable 
modifications of the standard test statistics can be deduced for 
a predefined non--zero source activity represented by 
a parameter $\beta$.
Resultant asymptotic significances are given by a transformation 
when $\alpha \to \alpha \beta$ is replaced in the standard 
significance formulas~\cite{Lim01}.
Further details of our derivation and its interesting features 
will be discussed elsewhere.
\begin{figure}[t]
\vspace{-2cm}
\centering
\includegraphics[width=0.48\textwidth]{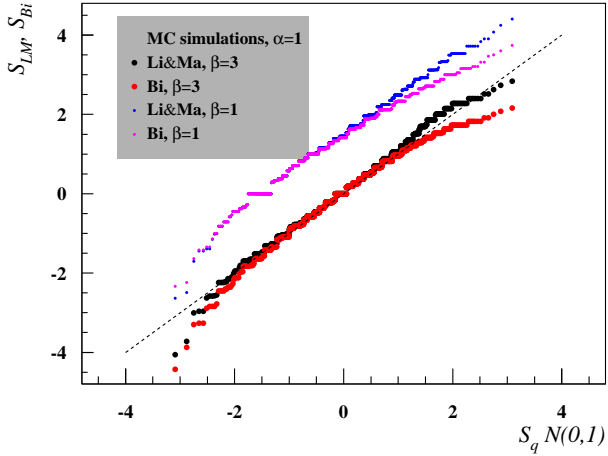}
\caption{Monte Carlo simulations.
We evaluate 1000 Poissonian events generated separately in the 
on-- and off--regions of equal weights ($\alpha = 1$) with means 
$\mon = 6.3$ and $\moff = 2.1$, respectively.
Resultant asymptotic significances are depicted in QQ plots as 
functions of Gaussian quantiles.
Li--Ma and binomial significances for the non--zero source
($\beta = 3$) are shown in black and red, respectively.
Small blue and magenta points represent Li--Ma and binomial 
significances for the null no--source hypothesis ($\beta = 1$) 
to be rejected.
The diagonal of the first quadrant is depicted by the dotted line.
}
\label{F01}
\end{figure}

\section{Time variability}
\label{Sec03}

\subsection{Monte Carlo simulations}
\label{Sec03a}

The statistical on--off tests were applied to the data 
simulated within the Monte Carlo (MC) method.
For this purpose, we generated 1000 Poissonian events separately 
in both on-- and off--regions of equal weights $\won = \woff$ 
giving $\alpha = 1$.
The reliability of our method was intentionally demonstrated 
on statistics based on small numbers.
The on-- and off--source means and intensities were 
$\mon = \lon = 6.3$ and $\moff = \loff = 2.1$, respectively.

First, we tested the MC data for the no--source hypothesis 
($\beta = 1$).
Our results are shown in \fig{F01} in a quantil--quantil 
plot (QQ plot).
Small blue and magenta points refer to asymptotic Li--Ma 
and binomial significances, respectively. 
Since these points do not lie on the indicated diagonal of the first
quadrant, the sample statistics $\SLM$ and $\SBii$ defined 
in~\app{App01} are shown not to come from Gaussian distribution 
with zero mean and unit variance.
In other words, the null no--source hypothesis ($\beta=1$) 
is demonstrated not to be true as expected.
Nonetheless, because the blue and magenta points in \fig{F01} 
approximately lie on a line, the distributions of the asymptotic 
Li--Ma and binomial statistics, $\SLM$ and $\SBii$, are to be 
linearly related to the standardized Gaussian distribution.

In the second step, we tested the non--zero source hypothesis 
assuming a parameter $\beta = 3$, i.e. $\mon = 3 \moff$ used 
as the input of MC simulations.
Resultant QQ plots for the asymptotic Li--Ma (big black points) 
and binomial (big red points) statistics are also depicted in \fig{F01}.
In this case, the relationship between studied sample statistics and
the standardized Gaussian distribution is well demonstrated.
These statistics lie on the diagonal of the first quadrant 
implying that $\SLM \sim \Noo(0,1)$ and $\SBii \sim \Noo(0,1)$ as well.

We conclude that the standard asymptotic measure of the level 
of significance of a source based on the null no--source 
conjecture can be trustworthy modified to search for an excess
or deficit of events with respect to a preassigned source strength 
using a non--zero source hypothesis.

\subsection{PKS~2155--304}
\label{Sec03b}

We have applied the modified on--off tests to demonstrate 
the change in the non--zero $\gamma$--activity of several 
observed sources.
In particular, here we present results showing to what extend 
the $\gamma$--ray variability observed in the direction of 
PKS~2155--304 on time scales of months and days can be verified 
on statistical grounds.
To this end, we have adopted experimental data collected by 
the H.E.S.S. telescopes in the 2002-2003 and 
2005-2007 campaigns~\cite{Hess01,Hess02}.

\begin{figure}[t!]
\vspace{-2cm}
\centering
\includegraphics[width=0.48\textwidth]{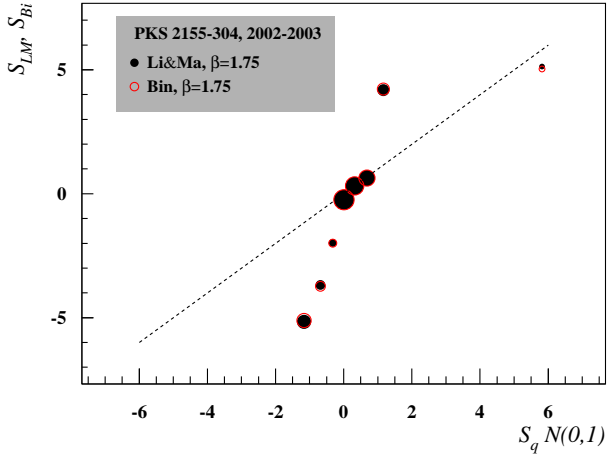}
\caption{
Asymptotic Li--Ma (black points) and binomial (red circles) 
significances for the non--zero $\gamma$--activity observed 
in the direction of  PKS~2155--304 are depicted in QQ plots 
as functions of Gaussian quantiles. 
All 8 sets of experimental counts observed in the 2002--2003 
H.E.S.S. campaign were taken from Table~3 in 
Ref.~\cite{Hess01}. 
Depicted significances were determined assuming a source 
parameter $\beta = 1.75$.
The observational time sequence of significances is visualized 
by increasing sizes of marks.
The diagonal of the first quadrant is shown by the dotted line.
}
\label{F02}
\end{figure}

Results of our analysis are presented in \figs{F02} and \figg{F03}.
In these figures, the distributions of asymptotic sample significances 
are compared to the standardized Gaussian distribution in QQ plots.
Both Li--Ma (black points) and binomial (red circles) significances
are shown.
Increasing sizes of marks refer to observational times of events 
for which asymptotic significances are depicted. 

In \fig{F02}, the results of our analysis of the experimental data 
collected in the period 2003--2005 by the H.E.S.S. instrument are shown.
In that time, 8 independent observations were recorded,
see Table~3 in Ref.~\cite{Hess01}.
For each of these events we calculated both asymptotic significances 
for the non--zero source.
The hypothetical $\gamma$--activity was characterized by a parameter 
$\beta = 1.75$.
This parameter roughly equals to an average observed on--source signal 
when compared to a background read out from the off--source 
region, i.e. $\beta \approx \langle \frac{\Non}{\alpha \Noff} \rangle$ 
where all quantities were taken from the aforementioned table in 
Ref.~\cite{Hess01}.
Except for one measurement at the end of 2002, the number 
of observed on-- or off--source counts ranged from several 
hundreds to over ten thousand.

The agreement between both studied statistics depicted in \fig{F02} 
is a salient feature.
It is well visible that their distributions are more dispersed 
than the reference standardized Gaussian distribution.
Except for the last three measurements with significances 
which are not inconsistent with the chosen average 
$\gamma$--activity of the source, the sample values of 
the remaining five significances, and its trend in particular, 
suggest that the collected data set is not drawn from 
the reference distribution.
This feature is to be interpreted as an unambiguous sign 
of monthly changes of the observed $\gamma$--ray flux 
from the investigated source.

Much more data on the PKS~2155--304 $\gamma$--activity has been 
collected in the 2005--2007 H.E.S.S. campaign~\cite{Hess02}.
In our analysis, only 29 events with MJD=53618--53705 and 
54264--54376 that were observed  before and after the huge 
flare have been included.
Using relevant data from Table~A.1 in Ref.~\cite{Hess02}, 
we obtained an average non--zero source activity expressed 
by a parameter $\beta = 1.38$.
In all cases, the number of observed on-- or off--source 
counts were above five hundred. 

In \fig{F03}, our results obtained from the later H.E.S.S. 
campaign are summarized.
Both studied asymptotic sample statistics, Li--Ma and binomial 
significances, agree with one another.
Their distributions are more dispersed than the standardized 
Gaussian distribution.
Moreover, their QQ plots are arced indicating that the sample 
significance distributions are more skewed than the reference 
distribution. 
This way, heavier tails observed in the sample distributions 
of the Li--Ma and binomial statistics suggest visible fluctuations
in the inspected data set.
Hence, not a few observations yielding the sample distributions 
of significances inconsistent with the reference distribution 
are to be considered as a signature of the time variability 
of the $\gamma$--ray flux observed from the studied source.

\section{Conclusions}
\label{Sec04}

We introduced the on--off tests with the null hypothesis that assumes 
a predefined source present in the on--source region.
Basic features of the Li--Ma as well as binomial asymptotic statistics 
with a given non--zero source were documented.
The MC results illustrate that the suggested method is satisfactory
even if statistics based on small numbers are examined.
The modified significance formulas were used to demonstrate the changes 
of the non--zero $\gamma$--ray flux observed by the H.E.S.S. telescopes 
in the direction of the PKS~2155--304 source on time scales of months 
and days.

\appendix

\section{The on--off method with non--zero source}
\label{App01}

We focus on a level of significance associated with
a statistical test trying to reject the null hypothesis stating that 
there is a source with a given activity in a suspected region.
We adopt notation used in $\gamma$--ray astronomy~\cite{Lim01}.
We deal with a hypothesis test in which the statistical significance 
of an excess or deficit of events with respect to a predefined 
source activity in a given region is established.

Let us assume that measured counts in the on--source region, $\Non$, 
come from a Poissonian distribution with a mean $\mon = \won \lon$, 
while off--source counts observed in the control off--source region, 
$\Noff$, come from a background Poissonian distribution with a mean 
$\moff = \woff \loff$.  
Here, $\lon$ and $\loff$ are unknown intensities of observed
on--source and off--source counts, i.e. $\Non \sim \Poo(\won \lon)$ and 
$\Noff \sim \Poo(\woff \loff)$.
Parameters $\won$ and $\woff$ assign known on--source and off--source 
weights, respectively. 

We will verify the null hypothesis stating $\lon = \beta \loff$, i.e.
$\mon = \won \lon = \alpha \beta \woff \loff = \alpha \beta \moff$,
where $\alpha = \frac{\won}{\woff}$ is the on--off parameter and 
a parameter $\beta > 0$ that is chosen in advance is responsible 
for an excess or deficit of events in the on--source region when 
compared to the off--source one if $\beta \not= 1$.
\begin{figure}[t!]
\vspace{-2cm}
\centering
\includegraphics[width=0.48\textwidth]{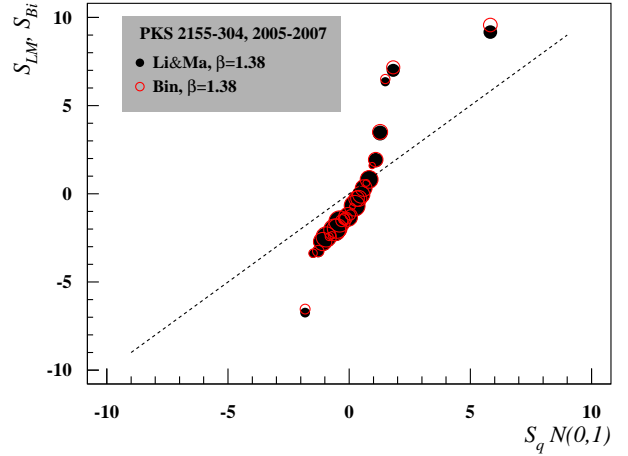}
\caption{
QQ plots of asymptotic Li--Ma (black points) and binomial 
(red circles) significances for the non--zero $\gamma$--activity 
of PKS 2155--304 are depicted.
Experimental data from the 2005--2007 H.E.S.S. campaign were taken 
from Table~A.1. in Ref.~\cite{Hess02}. 
Only 29 observations with MJD=53618--53705 and 54264--54376 away 
from the huge flare were included.
The null hypothesis was characterized by a source parameter 
$\beta = 1.38$.
For further details see caption to \fig{F02}.
}
\label{F03}
\end{figure}

\subsection{Binomial significance}
\label{App01a}

Let us assume a statistical test based on a conditional distribution.
Then, the probability that $\Non$ events is observed in 
the on--source region provided that $N = \Non + \Noff$ counts
measured in the whole inspected region 
follow the Poissonian distribution with the mean parameter 
$\mu = \mon + \moff$, is given by 
\beql{A01}
P(\Non \mid N ) = 
\frac{
P_{\Non}(\mon) 
P_{\Noff} (\moff) }
{P_{N}(\mu) } =
{N \choose \Non} 
q^{\Non} ( 1 - q )^{\Noff}.
\eeq
Here, $P_{k}(\mu) = \frac{\mu^{k}}{k!} e^{-\mu}$ assigns the Poissonian 
probability to observe $k$ events if the mean is $\mu$.
The number of on--source counts obeys the binomial distribution with 
the binomial parameter $q$. 
If $\lon = \beta \loff$ holds, then in addition  
\beql{A02}
q = q_{0} = \frac{\mon}{\mon + \moff} = 
\frac{\alpha \beta}{1 +\alpha \beta}.
\eeq
This way, the on--off problem is reduced to judge whether 
measured on--source counts $\Non$ can be viewed as a realization 
of the binomial distribution with the known parameter $q_{0}$ 
provided $N$ events are measured in the whole region 
under considerations.
The level of significance, the probability with which the 
null hypothesis ($\lon = \beta \loff$) is refused in favor of 
an excess of events in the on--source region ($\lon > \beta \loff$) 
if it is true (excess $p$--value), is 
\beql{A03}
\pex = 
\Ssum{k=\Non}{N}
{N \choose k} 
q_{0}^{k} ( 1 - q_{0} )^{N-k}.
\eeq
More precisely, the number of observed counts $\Non$ attains
a value the probability of which is under the stated hypothesis 
less than the predefined level of significance $\delta$, 
$\pex < \delta$. 
Accordingly, the level of significance with which the null hypothesis 
($\lon = \beta \loff$) is rejected in favor of a deficit of events 
in the on--source region, i.e. $\lon < \beta \loff$, if it is valid 
(deficit $p$--value) is
\beql{A04}
\pde = 
\Ssum{k=0}{\Non}
{N \choose k} 
q_{0}^{k} ( 1 - q_{0} )^{N-k}.
\eeq

If $\lon = \beta \loff$ and \eqa{A02} holds, then in the virtue of 
the de Moivre--Laplace theorem~\cite{Rao01} the binomial distribution 
of the number of on--source counts, $\Non \sim \Bii  (N, q_{0})$, 
is asymptotically Gaussian with the mean $N q_{0}$ and variance 
$N q_{0} ( 1- q_{0} )$ as $N$ tends to infinity. 
Inserting, on--off sample variables $\Non$ and $\Noff$, 
and the parameter $q_{0}$ given in \eqa{A02}, one recovers 
that the standardized sample variable 
\beql{A05}
\SBii = 
\frac{\Non - N q_{0}}{\sqrt{N q_{0} (1- q_{0})}} = 
\frac{\Non - \alpha \beta \Noff}{\sqrt{\alpha \beta ( \Non + \Noff )}},  
\eeq
can be considered asymptotically as drawn from the normal distribution 
with zero mean and unit variance, i.e. $\SBii \sim \Noo(0,1)$, 
expressing~\lq\lq $\mid \SBii \mid$ standard deviation result\rq\rq~for 
an excess ($\SBii > 0$) or deficit ($\SBii < 0$) of events, 
see e.g. Refs.~\cite{Lim01,Cou01}.
Needless to remind that the null no--source hypothesis test~\cite{Lim01} 
is regained when $\beta = 1$.

\subsection{Li--Ma significance}
\label{App01b}

Traditionally, the on--off problem is solved using asymptotic 
properties of the maximum likelihood ratio presented by 
T.Li and Y.Ma~\cite{Lim01}.
Also our extension for a non--zero source activity relies upon 
the ratio of the conditional likelihood functions of observed 
values of on-- and off--source counts given indicated 
parameters~\cite{Lim01}
\beql{B01}
R = \frac{L(\Non, \Noff \mid \mso, \hat \mbo)}
{L(\Non, \Noff \mid \hat \ms, \hat \mb)}.
\eeq
Here, $\mso$ is the mean of the source counts on the condition of 
the null hypothesis being true, i.e. $\mon = \alpha \beta \moff$, 
and $\hat \mbo$ is the conditional maximum likelihood estimate
of background counts given $\mso$. 
The maximum likelihood estimates of the means of source and 
background counts are assigned as $\hat \ms$ and $\hat \mb$.
Notice that the choice $\beta = 1$ implying $\mso = 0$ leads 
to the original Li--Ma problem~\cite{Lim01}.

On the general, the maximum likelihood estimates of the source 
and background means are~\cite{Lim01}
\beql{B02}
\hat \ms = \Non - \alpha \Noff, \qqb
\hat \mb = \alpha \Noff.
\eeq
The maximum likelihood estimates of the on-- and 
off--source Poissonian means are under the true null 
hypothesis, $\mon = \alpha \beta \moff$, given by 
(for more details on estimates of Poissonian parameters 
see e.g. Ref.~\cite{Rao01})
\beql{B03}
\hat \mono = \alpha \beta \hat \moffo =
\frac{\alpha \beta}{1 + \alpha \beta} (\Non + \Noff).
\eeq
Then one gets for the maximum likelihood estimate of 
the mean of background counts in the on--source region 
on the condition of the null hypothesis being true
\beql{B04}
\hat \mbo = \alpha \hat \moffo = 
\frac{\alpha}{1 + \alpha \beta} (\Non + \Noff).
\eeq

According to the theorem used in Ref.~\cite{Lim01}, if the present 
null hypothesis is valid, the statistics $\SLM^{2} = -2 \ln R$
will follow asymptotically ($\Non, \Noff \to \infty$) a 
$\chi^{2}$ distribution with one degree of freedom, 
$\SLM^{2} \sim \chi^{2}(1)$.
Therefore, if the null hypothesis is true, 
i.e. on-- and off--means satisfy $\mon = \alpha \beta \moff$ 
and counts in the on--source region are $\beta$--times larger than 
those expected from the background, then the non--negative square 
root of $\SLM^{2}$ has asymptotically the probability distribution as 
a random variable $\mid U \mid$, where $U \sim \Noo(0,1)$~\cite{Lim01}. 
Here, in agreement with the binomial asymptotics in \eqa{A05}, we 
consider the statistics $\SLM$ as non--negative if 
$\Non - \alpha \beta \Noff \ge 0$ and negative otherwise, i.e.
$\SLM = s \sqrt{\SLM^{2}}$ where $s = \sgn(\Non - \alpha \beta \Noff)$.
Hence, the modified random variable $\SLM$ can be assumed asymptotically 
as drawn from the standardized Gaussian distribution, 
$\SLM \sim \Noo(0,1)$.
Finally, using the estimates written in \eqb{B02}, \eqc{B03} 
and \eqc{B04}, this statistic now reads
\beql{B05}
\SLM = s \sqrt{2} 
\left\{ \Non \ln \Xon + \Noff \ln \Xoff \right\}^{\frac{1}{2}},
\eeq
where
$\Xon = 
\frac{(1+\alpha \beta)}{\alpha \beta} \frac{\Non}{\Non + \Noff}$,
and
$\Xoff =
(1+\alpha \beta) \frac{\Noff}{\Non + \Noff}$.
The value of the sample variable $\SLM$ is interpreted 
as the asymptotic significance of the observed result expressing
that a~\lq\lq $\mid \SLM \mid$ standard deviation event\rq\rq~above 
($\SLM > 0$) or below ($\SLM < 0$) the preassigned source 
activity represented by the source parameter $\beta$ is observed.
Choosing $\beta = 1$, the null no--source hypothesis is 
tested~\cite{Lim01}.

\vspace*{0.5cm}
\footnotesize{{\bf Acknowledgment:}
{This work was supported by the grants MSMT--CR LE13012, LG13007 and
MSM0021620859 of the Ministry of Education, Youth and Sports 
of the Czech Republic.
The authors would like to thank Petr Travnicek, Jakub Vicha and 
Jan Ebr for their help and support.}}



\end{document}